\documentclass[zpreprint,zbstpl,british]{zeus_paper}
\usepackage{tikz}
\usetikzlibrary{positioning}
\usetikzlibrary{shapes.misc} 
\usetikzlibrary{shapes.arrows}
\usetikzlibrary{arrows}
\RequirePackage{units}
\newif\ifzeusunit
\zeusunittrue

\newcommand{\Zdetdesc}{%
A detailed description of the ZEUS detector can be found 
elsewhere~\cite{zeus:1993:bluebook}. A brief outline of the 
components that are most relevant for this analysis is given
below.\xspace}

\newcommand{\Zsttdesc}[1]{%
The STT consisted of 48 sectors of two different sizes. Each sector
contained 192 (small sector) or 264 (large sector) straws of diameter
\unit[7.5]{\mm} arranged into 3 layers. The sectors were trapezoidal in shape
and each subtended an azimuthal angle of $60\degree$ -- 6 sectors
formed a so-called superlayer. A particle passing through the complete
detector traversed 8 superlayers, which were rotated around the beam
direction at angles of $30\degree$ or $15\degree$ to each other. The STT
covered the polar-angle region $5\degree < \theta < 23\degree$.
}
\newcommand{\Zcaldesc}{%
The high-resolution uranium--scintillator calorimeter (CAL)~\citeCAL
consisted of three parts: the forward (FCAL), the barrel (BCAL) and
the rear (RCAL) calorimeters. Each part was subdivided transversely
into towers and longitudinally into one electromagnetic section (EMC)
and either one (in RCAL) or two (in BCAL and FCAL) hadronic sections
(HAC). The smallest subdivision of the calorimeter was called a cell.
The CAL energy resolutions, as measured under test-beam conditions,
were $\sigma(E)/E=0.18/\sqrt{E}$ for electrons and
$\sigma(E)/E=0.35/\sqrt{E}$ for hadrons, with $E$ in \GeV.}

\newcommand{\ZlumidescA}[1]{%
The luminosity was measured using the Bethe--Heitler reaction
$ep\,\rightarrow\, e\gamma p$ by a luminosity detector which consisted
of independent lead--scintillator calorimeter\citePCAL and  magnetic
spectrometer\citeSPECTRO systems. The fractional systematic
uncertainty on the measured luminosity was \unit[#1]{\%}.}

\newcommand{\Zacknowledge}{%
We appreciate the contributions to the construction and maintenance of
the ZEUS detector of many people who are not listed as authors. The
HERA machine group and the DESY computing staff are especially
acknowledged for their success in providing excellent operation of the
collider and the data-analysis environment. We thank the DESY
directorate for their strong support and encouragement.}
\newcommand{\eVdist}{\kern-0.06667em}

\newcommand{\Mev}{{\text{Me}\eVdist\text{V\/}}}
\newcommand{\Gev}{{\text{Ge}\eVdist\text{V\/}}}

\newcommand{\gev}{{\,\text{Ge}\eVdist\text{V\/}}}

\newcommand{\pb}{\,\text{pb}}

\newcommand{\pbi}{\,\text{pb}^{-1}}
\newcommand{\fbi}{\,\text{fb}^{-1}}

\newcommand{\mm}{\,\text{mm}}
\newcommand{\cm}{\,\text{cm}}

\newcommand{\ns}{\,\text{ns}}

\newcommand{\rad}{\,\text{rad}}

\newcommand{\Tesla}{\,\text{T}}

\newcommand{\degree}{\ensuremath{^{\circ}}}

\newcommand{\GeV}{{\text{Ge}\eVdist\text{V\/}}}

\usepackage{lineno}
\chardef\usc=95
\chardef\til=126
\catcode`\@=11 
\DeclareRobustCommand\xdotspace{\futurelet\@let@token\@xdotspace}
\def\@xdotspace{%
  \ifx\@let@token.\else
  \ifx\@let@token\bgroup.\else
  \ifx\@let@token\egroup.\else
  \ifx\@let@token\/.\else
  \ifx\@let@token\ .\else
  \ifx\@let@token~.\else
  \ifx\@let@token!.\else
  \ifx\@let@token,.\else
  \ifx\@let@token:.\else
  \ifx\@let@token;.\else
  \ifx\@let@token?.\else
  \ifx\@let@token/.\else
  \ifx\@let@token'.\else
  \ifx\@let@token).\else
  \ifx\@let@token-.\else
  \ifx\@let@token\@xobeysp.\else
  \ifx\@let@token\space.\else
  \ifx\@let@token\@sptoken.\else
   .\space
   \fi\fi\fi\fi\fi\fi\fi\fi\fi\fi\fi\fi\fi\fi\fi\fi\fi\fi}
\catcode`\@=12 

\newcommand{\stru}[2]{%
   \relax\ifmmode\hbox{\vrule height#1 depth#2 width0pt}%
   \else\vrule height#1 depth#2 width0pt\fi}

\newcommand{\Ronum}[1]{\uppercase\expandafter{\romannumeral#1}}
\newcommand{\ronum}[1]{\expandafter{\romannumeral#1}}
\DeclareRobustCommand{\LaTeXZ}{%
  \LaTeX\kern-.05em4\kern-.1em
  {\raisebox{-0.2ex}{$\scriptstyle\text{ZEUS}$}}\xspace}

\newcommand{\slashfrac}[2]{%
  \raisebox{0.5ex}{\ensuremath #1}\kern-0.12em/\kern-0.08em
  \raisebox{-.8ex}{\ensuremath #2}}

\newcommand{\sqr}[3]{%
    {\vcenter{\hrule height.#3ex\hbox{\vrule width.#2ex height#1ex
     \kern#1ex\vrule width.#3ex}\hrule height.#2ex}}}

\catcode`\@=11 
\newcommand{\parenbar}{\mathpalette\p@renb@r}
\def\p@renb@r#1#2{\vbox{%
  \ifx#1\scriptscriptstyle \dimen@.7em\dimen@ii.2em\else
  \ifx#1\scriptstyle \dimen@.8em\dimen@ii.25em\else
  \dimen@1em\dimen@ii.4em\fi\fi \offinterlineskip
  \ialign{\hfill##\hfill\cr
    \vbox{\hrule width\dimen@ii}\cr
    \noalign{\vskip-.3ex}%
    \hbox to\dimen@{$\mathchar300\hfil\mathchar301$}\cr
    \noalign{\vskip-.3ex}%
    $#1#2$\cr}}}
\catcode`\@=12 

\ifzeusunit
  
\else
  
\fi

\newcommand{\IP}{{\rm I$\kern-0.01667em$P}\xspace}

\mathchardef\qsm=63
\mathchardef\pls=43
\mathchardef\mns=512
\mathchardef\plm=518
\mathchardef\eql=61
\mathchardef\smallleft=300
\mathchardef\smallright=301
\mathchardef\les=316
\mathchardef\gre=318
\mathchardef\leq=532
\mathchardef\grq=533

\catcode`\@=11 
\newcounter{pict@width}
\newcounter{pict@height}
\newlength{\pict@scale}
\newcommand{\psfigadd}[4]{%
\setcounter{pict@width}{1*\ratio{#2+\pict@scale/2}{\pict@scale}}
\setcounter{pict@height}{1*\ratio{#3+\pict@scale/2}{\pict@scale}}
\setlength{\unitlength}{\pict@scale}
\hbox to #2{\hspace{-\fill}\begin{picture}(\thepict@width,\thepict@height)
\put(0,0){\psfig{figure=#1,width=#2,height=#3,clip=}}
\SetScale{0.283466457}
\SetWidth{1.763889}
{#4}
\end{picture}}
}
\newcounter{pict@widthfst}
\newcounter{pict@widthscd}
\newcounter{pict@widthtot}
\newcommand{\psfigaddtwo}[7]{%
\setcounter{pict@widthfst}{1*\ratio{#2+\pict@scale/2}{\pict@scale}}
\setcounter{pict@widthscd}{1*\ratio{#2+#4+\pict@scale/2}{\pict@scale}}
\setcounter{pict@widthtot}{1*\ratio{#2+#4+#6+\pict@scale/2}{\pict@scale}}
\setcounter{pict@height}{1*\ratio{#3+\pict@scale/2}{\pict@scale}}
\setlength{\unitlength}{\pict@scale}
\hbox{\hspace{-\fill}\begin{picture}(\thepict@widthtot,\thepict@height)
\put(0,0){\psfig{figure=#1,width=#2,height=#3,clip=}}
\put(\thepict@widthscd,0){\psfig{figure=#5,width=#6,height=#3,clip=}}
\SetScale{0.283466457}
\SetWidth{1.763889}
{#7}
\end{picture}}
}
\newcommand{\psfigror}[4]{%
\setcounter{pict@width}{1*\ratio{#2+\pict@scale/2}{\pict@scale}}
\setcounter{pict@height}{1*\ratio{#3+\pict@scale/2}{\pict@scale}}
\setlength{\unitlength}{\pict@scale}
\hbox{\begin{picture}(\thepict@width,\thepict@height)
\put(0,\thepict@height){\psfig{figure=#1,width=#3,height=#2,clip=,angle=270}}
\SetScale{0.283466457}
\SetWidth{1.763889}
{#4}zeus_paper
\end{picture}}
}
\newcommand{\psfigrol}[4]{%
\setcounter{pict@width}{1*\ratio{#2+\pict@scale/2}{\pict@scale}}
\setcounter{pict@height}{1*\ratio{#3+\pict@scale/2}{\pict@scale}}
\setlength{\unitlength}{\pict@scale}
\hbox{\begin{picture}(\thepict@width,\thepict@height)
\put(0,0){\psfig{figure=#1,width=#3,height=#2,clip=,angle=90}}
\SetScale{0.283466457}
\SetWidth{1.763889}
{#4}
\end{picture}}
}
\catcode`\@=12 
\newlength\listtextwidth

\catcode`\@=11 
\newlength{\@tabfninsert}
\newlength{\@tabfnwidth}
\newcommand{\tabfootnote}[2]{%
  \setlength{\@tabfninsert}{0.8em}
  \setlength{\@tabfnwidth}{\textwidth}
  \addtolength{\@tabfnwidth}{-\@tabfninsert}
  \addtolength{\@tabfnwidth}{-0.4em}
  \noindent\makebox[\@tabfninsert][r]{\footnotesize$^{#1}$\hfil}\hfill%
  \parbox[t]{\@tabfnwidth}{\footnotesize #2\hfill}}
\catcode`\@=12 

\def\citeCTD{{\cite{%
nim:a279:290,*npps:b32:181,*nim:a338:254%
}}\xspace}
\def\citeMVD{{\cite{%
nim:a581:656%
}}\xspace}
\def\zeus_paperciteSTT{{\cite{%
nim:a535:191%
}}\xspace}
\def\citeCAL{{\cite{%
nim:a309:77,*nim:a309:101,*nim:a321:356,*nim:a336:23%
}}\xspace}

\def\citePCAL{{\cite{%
desy-92-066,*zfp:c63:391,*acpp:b32:2025%
}}\xspace}
\def\citeSPECTRO{{\cite{%
nim:a565:572%
}}\xspace}
\begin{document}
%
%
\prepnum{DESY--12--168}
\date{October 2012}
\title{Production of $\mathbf{Z^{0}}$ bosons in elastic and quasi-elastic 
\textit{\textbf{ep}} collisions at HERA 
}
                    
\author{ZEUS Collaboration}
\date{October 2012}

\abstract{\noindent
The production of $Z^{0}$ bosons in the reaction $ep\rightarrow eZ^{0}p^{\left(*\right)}$, where $p^{\left(*\right)}$ stands for a proton 
or a low-mass nucleon resonance, has been studied in $ep$ collisions at HERA using the ZEUS detector.  
The analysis is based on a data sample collected between 1996 and 2007, amounting to $\unit{496}{\pbi}$ of integrated luminosity.
The $Z^{0}$  was measured in the hadronic decay mode.
The elasticity of the events was ensured by a cut on  
$\eta_{{\rm max}} < 3.0$, where $\eta_{{\rm max}}$ is the maximum pseudorapidity of energy deposits in the calorimeter 
defined with respect to the proton beam direction.
A signal was observed at the $Z^{0}$  mass.
The cross section of the reaction $ep \rightarrow eZ^{0}p^{\left(*\right)}$ was measured to be 
$\sigma \left( ep \rightarrow eZ^{0}p^{\left(*\right)} \right) = \unit{ 0.13 \pm{0.06} \left( {\rm stat.} \right) \pm{0.01} \left( {\rm syst.} \right) }{\pb}$,
in agreement with the Standard Model prediction of $0.16 \pb$.
This is the first measurement of $Z^{0}$ production in $ep$ collisions.
}

\makezeustitle

\pagenumbering{Roman}

                                                   %
                                                   %
\begin{center}
{                      \Large  The ZEUS Collaboration              }
\end{center}

{\small


        {\raggedright
H.~Abramowicz$^{45, ah}$, 
I.~Abt$^{35}$, 
L.~Adamczyk$^{13}$, 
M.~Adamus$^{54}$, 
R.~Aggarwal$^{7, c}$, 
S.~Antonelli$^{4}$, 
P.~Antonioli$^{3}$, 
A.~Antonov$^{33}$, 
M.~Arneodo$^{50}$, 
O.~Arslan$^{5}$
% ...
        }

\newpage


\makebox[3em]{$^{1}$}
\begin{minipage}[t]{14cm}
{\it Argonne National Laboratory, Argonne, Illinois 60439-4815, USA}~$^{A}$

\end{minipage}\\
\makebox[3em]{$^{2}$}
\begin{minipage}[t]{14cm}
{\it Andrews University, Berrien Springs, Michigan 49104-0380, USA}

\end{minipage}\\
\makebox[3em]{$^{3}$}
\begin{minipage}[t]{14cm}
{\it INFN Bologna, Bologna, Italy}~$^{B}$

\end{minipage}\\
\makebox[3em]{$^{4}$}
\begin{minipage}[t]{14cm}
{\it University and INFN Bologna, Bologna, Italy}~$^{B}$

\end{minipage}\\
\makebox[3em]{$^{5}$}
\begin{minipage}[t]{14cm}
{\it Physikalisches Institut der Universit\"at Bonn,
Bonn, Germany}~$^{C}$

\end{minipage}\\
\makebox[3em]{$^{6}$}
\begin{minipage}[t]{14cm}
{\it H.H.~Wills Physics Laboratory, University of Bristol,
Bristol, United Kingdom}~$^{D}$

\end{minipage}\\
\makebox[3em]{$^{7}$}
\begin{minipage}[t]{14cm}
{\it Panjab University, Department of Physics, Chandigarh, India}

\end{minipage}\\
\makebox[3em]{$^{8}$}
\begin{minipage}[t]{14cm}
{\it Calabria University,
Physics Department and INFN, Cosenza, Italy}~$^{B}$

\end{minipage}\\
\makebox[3em]{$^{9}$}
\begin{minipage}[t]{14cm}
{\it Institute for Universe and Elementary Particles, Chonnam National University,\\
Kwangju, South Korea}

\end{minipage}\\
\makebox[3em]{$^{10}$}
\begin{minipage}[t]{14cm}
{\it Jabatan Fizik, Universiti Malaya, 50603 Kuala Lumpur, Malaysia}~$^{E}$

\end{minipage}\\
\makebox[3em]{$^{11}$}
\begin{minipage}[t]{14cm}
{\it Nevis Laboratories, Columbia University, Irvington on Hudson,
New York 10027, USA}~$^{F}$

\end{minipage}\\
\makebox[3em]{$^{12}$}
\begin{minipage}[t]{14cm}
{\it The Henryk Niewodniczanski Institute of Nuclear Physics, Polish Academy of \\
Sciences, Krakow, Poland}~$^{G}$

\end{minipage}\\
\makebox[3em]{$^{13}$}
\begin{minipage}[t]{14cm}
{\it AGH-University of Science and Technology, Faculty of Physics and Applied Computer
Science, Krakow, Poland}~$^{H}$

\end{minipage}\\
\makebox[3em]{$^{14}$}
\begin{minipage}[t]{14cm}
{\it Department of Physics, Jagellonian University, Cracow, Poland}

\end{minipage}\\
\makebox[3em]{$^{15}$}
\begin{minipage}[t]{14cm}
{\it Deutsches Elektronen-Synchrotron DESY, Hamburg, Germany}

\end{minipage}\\
\makebox[3em]{$^{16}$}
\begin{minipage}[t]{14cm}
{\it Deutsches Elektronen-Synchrotron DESY, Zeuthen, Germany}

\end{minipage}\\
\makebox[3em]{$^{17}$}
\begin{minipage}[t]{14cm}
{\it INFN Florence, Florence, Italy}~$^{B}$

\end{minipage}\\
\makebox[3em]{$^{18}$}
\begin{minipage}[t]{14cm}
{\it University and INFN Florence, Florence, Italy}~$^{B}$

\end{minipage}\\
\makebox[3em]{$^{19}$}
\begin{minipage}[t]{14cm}
{\it Fakult\"at f\"ur Physik der Universit\"at Freiburg i.Br.,
Freiburg i.Br., Germany}

\end{minipage}\\
\makebox[3em]{$^{20}$}
\begin{minipage}[t]{14cm}
{\it School of Physics and Astronomy, University of Glasgow,
Glasgow, United Kingdom}~$^{D}$

\end{minipage}\\
\makebox[3em]{$^{21}$}
\begin{minipage}[t]{14cm}
{\it Department of Engineering in Management and Finance, Univ. of
the Aegean, Chios, Greece}

\end{minipage}\\
\makebox[3em]{$^{22}$}
\begin{minipage}[t]{14cm}
{\it Hamburg University, Institute of Experimental Physics, Hamburg,
Germany}~$^{I}$

\end{minipage}\\
\makebox[3em]{$^{23}$}
\begin{minipage}[t]{14cm}
{\it Imperial College London, High Energy Nuclear Physics Group,
London, United Kingdom}~$^{D}$

\end{minipage}\\
\makebox[3em]{$^{24}$}
\begin{minipage}[t]{14cm}
{\it Institute of Particle and Nuclear Studies, KEK,
Tsukuba, Japan}~$^{J}$

\end{minipage}\\
\makebox[3em]{$^{25}$}
\begin{minipage}[t]{14cm}
{\it Institute of Physics and Technology of Ministry of Education and
Science of Kazakhstan, Almaty, Kazakhstan}

\end{minipage}\\
\makebox[3em]{$^{26}$}
\begin{minipage}[t]{14cm}
{\it Institute for Nuclear Research, National Academy of Sciences, Kyiv, Ukraine}

\end{minipage}\\
\makebox[3em]{$^{27}$}
\begin{minipage}[t]{14cm}
{\it Department of Nuclear Physics, National Taras Shevchenko University of Kyiv, Kyiv, Ukraine}

\end{minipage}\\
\makebox[3em]{$^{28}$}
\begin{minipage}[t]{14cm}
{\it Kyungpook National University, Center for High Energy Physics, Daegu,
South Korea}~$^{K}$

\end{minipage}\\
\makebox[3em]{$^{29}$}
\begin{minipage}[t]{14cm}
{\it Institut de Physique Nucl\'{e}aire, Universit\'{e} Catholique de Louvain, Louvain-la-Neuve,\\
Belgium}~$^{L}$

\end{minipage}\\
\makebox[3em]{$^{30}$}
\begin{minipage}[t]{14cm}
{\it Departamento de F\'{\i}sica Te\'orica, Universidad Aut\'onoma
de Madrid, Madrid, Spain}~$^{M}$

\end{minipage}\\
\makebox[3em]{$^{31}$}
\begin{minipage}[t]{14cm}
{\it Department of Physics, McGill University,
Montr\'eal, Qu\'ebec, Canada H3A 2T8}~$^{N}$

\end{minipage}\\
\makebox[3em]{$^{32}$}
\begin{minipage}[t]{14cm}
{\it Meiji Gakuin University, Faculty of General Education,
Yokohama, Japan}~$^{J}$

\end{minipage}\\
\makebox[3em]{$^{33}$}
\begin{minipage}[t]{14cm}
{\it Moscow Engineering Physics Institute, Moscow, Russia}~$^{O}$

\end{minipage}\\
\makebox[3em]{$^{34}$}
\begin{minipage}[t]{14cm}
{\it Lomonosov Moscow State University, Skobeltsyn Institute of Nuclear Physics,
Moscow, Russia}~$^{P}$

\end{minipage}\\
\makebox[3em]{$^{35}$}
\begin{minipage}[t]{14cm}
{\it Max-Planck-Institut f\"ur Physik, M\"unchen, Germany}

\end{minipage}\\
\makebox[3em]{$^{36}$}
\begin{minipage}[t]{14cm}
{\it NIKHEF and University of Amsterdam, Amsterdam, Netherlands}~$^{Q}$

\end{minipage}\\
\makebox[3em]{$^{37}$}
\begin{minipage}[t]{14cm}
{\it Physics Department, Ohio State University,
Columbus, Ohio 43210, USA}~$^{A}$

\end{minipage}\\
\makebox[3em]{$^{38}$}
\begin{minipage}[t]{14cm}
{\it Department of Physics, University of Oxford,
Oxford, United Kingdom}~$^{D}$

\end{minipage}\\
\makebox[3em]{$^{39}$}
\begin{minipage}[t]{14cm}
{\it INFN Padova, Padova, Italy}~$^{B}$

\end{minipage}\\
\makebox[3em]{$^{40}$}
\begin{minipage}[t]{14cm}
{\it Dipartimento di Fisica dell' Universit\`a and INFN,
Padova, Italy}~$^{B}$

\end{minipage}\\
\makebox[3em]{$^{41}$}
\begin{minipage}[t]{14cm}
{\it Department of Physics, Pennsylvania State University, University Park,\\
Pennsylvania 16802, USA}~$^{F}$

\end{minipage}\\
\makebox[3em]{$^{42}$}
\begin{minipage}[t]{14cm}
{\it Polytechnic University, Tokyo, Japan}~$^{J}$

\end{minipage}\\
\makebox[3em]{$^{43}$}
\begin{minipage}[t]{14cm}
{\it Dipartimento di Fisica, Universit\`a 'La Sapienza' and INFN,
Rome, Italy}~$^{B}$

\end{minipage}\\
\makebox[3em]{$^{44}$}
\begin{minipage}[t]{14cm}
{\it Rutherford Appleton Laboratory, Chilton, Didcot, Oxon,
United Kingdom}~$^{D}$

\end{minipage}\\
\makebox[3em]{$^{45}$}
\begin{minipage}[t]{14cm}
{\it Raymond and Beverly Sackler Faculty of Exact Sciences, School of Physics, \\
Tel Aviv University, Tel Aviv, Israel}~$^{R}$

\end{minipage}\\
\makebox[3em]{$^{46}$}
\begin{minipage}[t]{14cm}
{\it Department of Physics, Tokyo Institute of Technology,
Tokyo, Japan}~$^{J}$

\end{minipage}\\
\makebox[3em]{$^{47}$}
\begin{minipage}[t]{14cm}
{\it Department of Physics, University of Tokyo,
Tokyo, Japan}~$^{J}$

\end{minipage}\\
\makebox[3em]{$^{48}$}
\begin{minipage}[t]{14cm}
{\it Tokyo Metropolitan University, Department of Physics,
Tokyo, Japan}~$^{J}$

\end{minipage}\\
\makebox[3em]{$^{49}$}
\begin{minipage}[t]{14cm}
{\it Universit\`a di Torino and INFN, Torino, Italy}~$^{B}$

\end{minipage}\\
\makebox[3em]{$^{50}$}
\begin{minipage}[t]{14cm}
{\it Universit\`a del Piemonte Orientale, Novara, and INFN, Torino,
Italy}~$^{B}$

\end{minipage}\\
\makebox[3em]{$^{51}$}
\begin{minipage}[t]{14cm}
{\it Department of Physics, University of Toronto, Toronto, Ontario,
Canada M5S 1A7}~$^{N}$

\end{minipage}\\
\makebox[3em]{$^{52}$}
\begin{minipage}[t]{14cm}
{\it Physics and Astronomy Department, University College London,
London, United Kingdom}~$^{D}$

\end{minipage}\\
\makebox[3em]{$^{53}$}
\begin{minipage}[t]{14cm}
{\it Faculty of Physics, University of Warsaw, Warsaw, Poland}

\end{minipage}\\
\makebox[3em]{$^{54}$}
\begin{minipage}[t]{14cm}
{\it National Centre for Nuclear Research, Warsaw, Poland}

\end{minipage}\\
\makebox[3em]{$^{55}$}
\begin{minipage}[t]{14cm}
{\it Department of Particle Physics and Astrophysics, Weizmann
Institute, Rehovot, Israel}

\end{minipage}\\
\makebox[3em]{$^{56}$}
\begin{minipage}[t]{14cm}
{\it Department of Physics, University of Wisconsin, Madison,
Wisconsin 53706, USA}~$^{A}$

\end{minipage}\\
\makebox[3em]{$^{57}$}
\begin{minipage}[t]{14cm}
{\it Department of Physics, York University, Ontario, Canada M3J 1P3}~$^{N}$

\end{minipage}\\
\vspace{30em} \pagebreak[4]


\makebox[3ex]{$^{ A}$}
\begin{minipage}[t]{14cm}
 supported by the US Department of Energy\
\end{minipage}\\
\makebox[3ex]{$^{ B}$}
\begin{minipage}[t]{14cm}
 supported by the Italian National Institute for Nuclear Physics (INFN) \
\end{minipage}\\
\makebox[3ex]{$^{ C}$}
\begin{minipage}[t]{14cm}
 supported by the German Federal Ministry for Education and Research (BMBF), under
 contract No. 05 H09PDF\
\end{minipage}\\
\makebox[3ex]{$^{ D}$}
\begin{minipage}[t]{14cm}
 supported by the Science and Technology Facilities Council, UK\
\end{minipage}\\
\makebox[3ex]{$^{ E}$}
\begin{minipage}[t]{14cm}
 supported by HIR and UMRG grants from Universiti Malaya, and an ERGS grant from the
 Malaysian Ministry for Higher Education\
\end{minipage}\\
\makebox[3ex]{$^{ F}$}
\begin{minipage}[t]{14cm}
 supported by the US National Science Foundation. Any opinion,
 findings and conclusions or recommendations expressed in this material
 are those of the authors and do not necessarily reflect the views of the
 National Science Foundation.\
\end{minipage}\\
\makebox[3ex]{$^{ G}$}
\begin{minipage}[t]{14cm}
 supported by the Polish Ministry of Science and Higher Education as a scientific project No.
 DPN/N188/DESY/2009\
\end{minipage}\\
\makebox[3ex]{$^{ H}$}
\begin{minipage}[t]{14cm}
 supported by the Polish Ministry of Science and Higher Education and its grants
 for Scientific Research\
\end{minipage}\\
\makebox[3ex]{$^{ I}$}
\begin{minipage}[t]{14cm}
 supported by the German Federal Ministry for Education and Research (BMBF), under
 contract No. 05h09GUF, and the SFB 676 of the Deutsche Forschungsgemeinschaft (DFG) \
\end{minipage}\\
\makebox[3ex]{$^{ J}$}
\begin{minipage}[t]{14cm}
 supported by the Japanese Ministry of Education, Culture, Sports, Science and Technology
 (MEXT) and its grants for Scientific Research\
\end{minipage}\\
\makebox[3ex]{$^{ K}$}
\begin{minipage}[t]{14cm}
 supported by the Korean Ministry of Education and Korea Science and Engineering
 Foundation\
\end{minipage}\\
\makebox[3ex]{$^{ L}$}
\begin{minipage}[t]{14cm}
 supported by FNRS and its associated funds (IISN and FRIA) and by an Inter-University
 Attraction Poles Programme subsidised by the Belgian Federal Science Policy Office\
\end{minipage}\\
\makebox[3ex]{$^{ M}$}
\begin{minipage}[t]{14cm}
 supported by the Spanish Ministry of Education and Science through funds provided by
 CICYT\
\end{minipage}\\
\makebox[3ex]{$^{ N}$}
\begin{minipage}[t]{14cm}
 supported by the Natural Sciences and Engineering Research Council of Canada (NSERC) \
\end{minipage}\\
\makebox[3ex]{$^{ O}$}
\begin{minipage}[t]{14cm}
 partially supported by the German Federal Ministry for Education and Research (BMBF)\
\end{minipage}\\
\makebox[3ex]{$^{ P}$}
\begin{minipage}[t]{14cm}
 supported by RF Presidential grant N 3920.2012.2 for the Leading Scientific Schools and by
 the Russian Ministry of Education and Science through its grant for Scientific Research on
 High Energy Physics\
\end{minipage}\\
\makebox[3ex]{$^{ Q}$}
\begin{minipage}[t]{14cm}
 supported by the Netherlands Foundation for Research on Matter (FOM)\
\end{minipage}\\
\makebox[3ex]{$^{ R}$}
\begin{minipage}[t]{14cm}
 supported by the Israel Science Foundation\
\end{minipage}\\
\vspace{30em} \pagebreak[4]


\makebox[3ex]{$^{ a}$}
\begin{minipage}[t]{14cm}
now at University of Salerno, Italy\
\end{minipage}\\
\makebox[3ex]{$^{ b}$}
\begin{minipage}[t]{14cm}
now at Queen Mary University of London, United Kingdom\
\end{minipage}\\
\makebox[3ex]{$^{ c}$}
\begin{minipage}[t]{14cm}
also funded by Max Planck Institute for Physics, Munich, Germany\
\end{minipage}\\
\makebox[3ex]{$^{ d}$}
\begin{minipage}[t]{14cm}
also Senior Alexander von Humboldt Research Fellow at Hamburg University,
 Institute of Experimental Physics, Hamburg, Germany\
\end{minipage}\\
\makebox[3ex]{$^{ e}$}
\begin{minipage}[t]{14cm}
also at Cracow University of Technology, Faculty of Physics,
 Mathemathics and Applied Computer Science, Poland\
\end{minipage}\\
\makebox[3ex]{$^{ f}$}
\begin{minipage}[t]{14cm}
supported by the research grant No. 1 P03B 04529 (2005-2008)\
\end{minipage}\\
\makebox[3ex]{$^{ g}$}
\begin{minipage}[t]{14cm}
supported by the Polish National Science Centre, project No. DEC-2011/01/BST2/03643\
\end{minipage}\\
\makebox[3ex]{$^{ h}$}
\begin{minipage}[t]{14cm}
now at Rockefeller University, New York, NY
 10065, USA\
\end{minipage}\\
\makebox[3ex]{$^{ i}$}
\begin{minipage}[t]{14cm}
now at DESY group FS-CFEL-1\
\end{minipage}\\
\makebox[3ex]{$^{ j}$}
\begin{minipage}[t]{14cm}
now at Institute of High Energy Physics, Beijing, China\
\end{minipage}\\
\makebox[3ex]{$^{ k}$}
\begin{minipage}[t]{14cm}
now at DESY group FEB, Hamburg, Germany\
\end{minipage}\\
\makebox[3ex]{$^{ l}$}
\begin{minipage}[t]{14cm}
also at Moscow State University, Russia\
\end{minipage}\\
\makebox[3ex]{$^{ m}$}
\begin{minipage}[t]{14cm}
now at University of Liverpool, United Kingdom\
\end{minipage}\\
\makebox[3ex]{$^{ n}$}
\begin{minipage}[t]{14cm}
now at CERN, Geneva, Switzerland\
\end{minipage}\\
\makebox[3ex]{$^{ o}$}
\begin{minipage}[t]{14cm}
also affiliated with Universtiy College London, UK\
\end{minipage}\\
\makebox[3ex]{$^{ p}$}
\begin{minipage}[t]{14cm}
now at Goldman Sachs, London, UK\
\end{minipage}\\
\makebox[3ex]{$^{ q}$}
\begin{minipage}[t]{14cm}
also at Institute of Theoretical and Experimental Physics, Moscow, Russia\
\end{minipage}\\
\makebox[3ex]{$^{ r}$}
\begin{minipage}[t]{14cm}
also at FPACS, AGH-UST, Cracow, Poland\
\end{minipage}\\
\makebox[3ex]{$^{ s}$}
\begin{minipage}[t]{14cm}
partially supported by Warsaw University, Poland\
\end{minipage}\\
\makebox[3ex]{$^{ t}$}
\begin{minipage}[t]{14cm}
supported by the Alexander von Humboldt Foundation\
\end{minipage}\\
\makebox[3ex]{$^{ u}$}
\begin{minipage}[t]{14cm}
now at Istituto Nucleare di Fisica Nazionale (INFN), Pisa, Italy\
\end{minipage}\\
\makebox[3ex]{$^{ v}$}
\begin{minipage}[t]{14cm}
now at Haase Energie Technik AG, Neum\"unster, Germany\
\end{minipage}\\
\makebox[3ex]{$^{ w}$}
\begin{minipage}[t]{14cm}
now at Department of Physics, University of Bonn, Germany\
\end{minipage}\\
\makebox[3ex]{$^{ x}$}
\begin{minipage}[t]{14cm}
also affiliated with DESY, Germany\
\end{minipage}\\
\makebox[3ex]{$^{ y}$}
\begin{minipage}[t]{14cm}
also at University of Tokyo, Japan\
\end{minipage}\\
\makebox[3ex]{$^{ z}$}
\begin{minipage}[t]{14cm}
now at Kobe University, Japan\
\end{minipage}\\
\makebox[3ex]{$^{\dagger}$}
\begin{minipage}[t]{14cm}
 deceased \
\end{minipage}\\
\makebox[3ex]{$^{aa}$}
\begin{minipage}[t]{14cm}
supported by DESY, Germany\
\end{minipage}\\
\makebox[3ex]{$^{ab}$}
\begin{minipage}[t]{14cm}
member of National Technical University of Ukraine, Kyiv Polytechnic Institute,
 Kyiv, Ukraine\
\end{minipage}\\
\makebox[3ex]{$^{ac}$}
\begin{minipage}[t]{14cm}
member of National University of Kyiv - Mohyla Academy, Kyiv, Ukraine\
\end{minipage}\\
\makebox[3ex]{$^{ad}$}
\begin{minipage}[t]{14cm}
Alexander von Humboldt Professor; also at DESY and University of Oxford\
\end{minipage}\\
\makebox[3ex]{$^{ae}$}
\begin{minipage}[t]{14cm}
STFC Advanced Fellow\
\end{minipage}\\
\makebox[3ex]{$^{af}$}
\begin{minipage}[t]{14cm}
now at LNF, Frascati, Italy\
\end{minipage}\\
\makebox[3ex]{$^{ag}$}
\begin{minipage}[t]{14cm}
This material was based on work supported by the
 National Science Foundation, while working at the Foundation.\
\end{minipage}\\
\makebox[3ex]{$^{ah}$}
\begin{minipage}[t]{14cm}
also at Max Planck Institute for Physics, Munich, Germany, External Scientific Member\
\end{minipage}\\
\makebox[3ex]{$^{ai}$}
\begin{minipage}[t]{14cm}
now at Tokyo Metropolitan University, Japan\
\end{minipage}\\
\makebox[3ex]{$^{aj}$}
\begin{minipage}[t]{14cm}
now at Nihon Institute of Medical Science, Japan\
\end{minipage}\\
\makebox[3ex]{$^{ak}$}
\begin{minipage}[t]{14cm}
now at Osaka University, Osaka, Japan\
\end{minipage}\\
\makebox[3ex]{$^{al}$}
\begin{minipage}[t]{14cm}
also at \L\'{o}d\'{z} University, Poland\
\end{minipage}\\
\makebox[3ex]{$^{am}$}
\begin{minipage}[t]{14cm}
member of \L\'{o}d\'{z} University, Poland\
\end{minipage}\\
\makebox[3ex]{$^{an}$}
\begin{minipage}[t]{14cm}
now at Department of Physics, Stockholm University, Stockholm, Sweden\
\end{minipage}\\
\makebox[3ex]{$^{ao}$}
\begin{minipage}[t]{14cm}
also at Cardinal Stefan Wyszy\'nski University, Warsaw, Poland\
\end{minipage}\\

}

\newpage

\pagenumbering{arabic}
%
%
\section{Introduction}
\label{sec-int}
The production of electroweak bosons in $ep$ collisions is a good benchmark process for testing the Standard Model\,(SM).
Even though the expected numbers of events
for $W^{\pm}$ and $Z^{0}$ production are low, the measurement of the cross sections 
of these processes is important
as some extensions of the SM predict anomalous couplings and thus changes in
these cross sections.
A measurement of the cross section for $W^{\pm}$ production at HERA has been performed by H1 and ZEUS\,\cite{Wpaper} in events containing an isolated lepton and missing transverse momentum, 
giving a cross section
$\sigma \left(ep\rightarrow W^{\pm}X \right) = \unit{ 1.06 \pm 0.17 \left({\rm stat.} \oplus {\rm  syst.}\right) }{\pb},$ in good agreement with the SM prediction. 
The cross section for $Z^{0}$ production is predicted to be $\unit{0.4}{\pb}$.

This paper reports on a measurement of the production of $Z^{0}$  bosons in $e^{\pm}p$  collisions using an integrated luminosity 
of about $\unit{0.5}{\fbi}$.
The hadronic decay mode was chosen\footnote{The $Z^{0}$ decay into charged lepton pairs was studied
in a previous ZEUS publication\,\cite{phy.lett:b680:13-23}.} because of its large branching ratio and because it 
allows the excellent resolution of the ZEUS hadronic calorimeter to be exploited to the full.
The analysis was restricted to elastic and quasi-elastic $Z^{0}$  production in order to suppress QCD multi-jet background. 
The selected process is $ep \rightarrow eZ^{0}p^{\left( * \right)}$, where 
$p^{\left(*\right)}$ stands for a proton (elastic process) or a low-mass nucleon resonance (quasi-elastic process).

\begin{figure}[p]
\vfill
\begin{center}
\includegraphics[angle=-90, width=0.5\hsize]{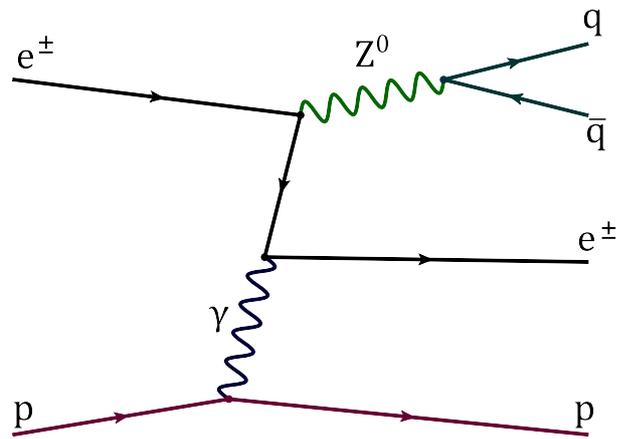}
\end{center}
\caption{Example of a leading-order diagram of $Z^{0}$ boson production and subsequent hadronic decay (into quark $q$ and antiquark $\bar{q}$) 
in  $ep\rightarrow eZ^{0}p$. }
\label{diagram}
\vfill
\end{figure}

Figure \ref{diagram} shows a leading-order (LO) diagram of $Z^{0}$ production
with subsequent hadronic decay.
In such events, there are at least two hadronic jets with high transverse energies, and no hadronic energy deposits around the 
forward\footnote{The ZEUS coordinate system 
is a right-handed Cartesian system, with the $Z$ axis pointing in the proton beam direction, referred to as the forward direction, 
and the $X$ axis pointing towards the centre of HERA.
The coordinate origin is at the nominal interaction point. The pseudorapidity is defined as 
$\eta = -\ln\left( \tan{\frac{\theta}{2}} \right)$, where the polar angle, $\theta$, 
is measured with respect to the proton beam direction.} 
direction, in contrast to what would be expected in inelastic collisions.

\section{Experimental set-up}
\label{sec-exp}
HERA was the world's only high-energy $ep$ collider, with an electron\footnote{The term electron also refers to positrons if not stated otherwise.} 
beam of $\unit{27.6\,}{\Gev}$ and a proton beam of 
$\unit{920\,}{\Gev}$\,($\unit{820\,}{\Gev}$ until 1997).
For this analysis, $e^{\pm}p$ collision data collected with the ZEUS detector between 1996 and 2007, 
amounting to $\unit{496}{\pbi}$ of integrated luminosity, have been used.
They consist of $\unit{289}{\pbi}$ of $e^{+}p$  data and $\unit{207}{\pbi}$ of $e^{-}p$ data.

After 2003, HERA was operated with a polarised lepton beam.
When combining the data taken with negative and positive polarisations,
the average polarisation is less than 1\% and its effect was neglected in this analysis.

\Zdetdesc

Charged particles were tracked
in the central tracking detector (CTD)~\citeCTD, which operated in a magnetic field of $1.43\Tesla$ provided by a thin superconducting solenoid.
The CTD consisted of 72~cylindrical drift chamber layers, 
organised in nine superlayers covering the polar-angle region \mbox{$15^\circ<\theta<164^\circ$}. 
For the data taken after 2001, the CTD was complemented by a silicon microvertex
detector (MVD)~\citeMVD, consisting of three active layers in the barrel and four disks in the forward region.

\Zcaldesc



\ZlumidescA{2}

\section{Monte Carlo simulations}\label{mcsimulation}
Monte Carlo (MC) simulations were made to simulate the $Z^{0}$ production process.
They were used to correct for instrumental effects and selection acceptance 
and to provide a template for the shape of the invariant-mass distribution of the $Z^{0}$  signal.
The EPVEC program\cite{np:b375:3} was used to generate the signal events at the parton level. 
The following $Z^{0}$ production processes are considered in EPVEC:
\begin{itemize}
\item elastic scattering, $ep \rightarrow eZ^{0}p$, where the proton stays intact;
\item quasi-elastic scattering,  $ep \rightarrow eZ^{0}p^{*}$, where the proton is transformed into a nucleon resonance $p^{*}$;
\item deep inelastic scattering\,(DIS), $\gamma^{*}p \rightarrow Z^{0}X$, in the region $Q^{2} > \unit{4\,}{\Gev^{2}}$, 
where $Q^{2}$ is the virtuality of the photon exchanged between the electron and proton;
\item resolved photoproduction, $\gamma p \rightarrow \left( q\bar{q} \rightarrow Z^{0} \right)X$, where one of the quarks is 
a constituent of the resolved photon and the other quark is a constituent of the proton.
\end{itemize}
In EPVEC the first two processes 
are calculated using form factors and structure functions fitted directly to experimental data. 
Note that, even if the virtuality of the exchanged photon is small, the scattered electron could receive a large momentum transfer when the 
$Z^{0}$ is radiated from the lepton line.
In the last two processes, the proton breaks up.
The DIS process is calculated in the quark--parton model using a full set of leading-order Feynman diagrams.
Resolved photoproduction is parametrised using a photon structure function and is carefully matched to the DIS region.
The cross section of $Z^{0}$ production is calculated to be $\unit{0.16}{\pb}$ for elastic and quasi-elastic processes and $\unit{0.24}{\pb}$ for DIS and resolved photoproduction.
The difference between $e^{+}p$ and $e^{-}p$ cross sections is negligible for this analysis (<1\% for the DIS process).
A correction, based on the MC cross section,  was made to account for the part of data taken at the centre-of-mass energy $\sqrt{s}=\unit{300\,}{\Gev}$, 
so that the result is quoted at $\sqrt{s}=\unit{318\,}{\Gev}$.

After the parton-level generation by EPVEC, PYTHIA 5.6\cite{manual:cern-th-6488/92} was used to simulate initial- and final-state parton showers with the fragmentation into hadrons using the Lund string model\cite{prep:97:31} as implemented in JETSET 7.3\cite{manual:cern-th-6488/92}.
The generated MC events were passed through the ZEUS detector
and trigger simulation programs based on GEANT 3.13\cite{tech:cern-dd-ee-84-1}.
They were reconstructed and analysed by the same programs as the data.

A reliable prediction of background events with the signal topology, which are predominantly due to the diffractive photoproduction 
of jets of high transverse momentum, 
is currently not available.
Therefore, the background shape of the invariant-mass distribution was estimated with a data-driven method.
The normalisation was determined by a fit to the data.

\section{Event reconstruction and selection}
\label{obj_recon}

The events used in this analysis were selected online by the ZEUS three-level trigger system\cite{uproc:chep:1992:222}, using a combination of several trigger chains which 
were mainly based on requirements of large transverse energy deposited in the calorimeter.
In the offline selection, further criteria were imposed in order to separate the signal from the background.

The events are characterised by the presence of at least two jets of high transverse energy
and, for a fraction of events, by the presence of a reconstructed scattered electron.
In order to select events with a $Z^{0}$ decaying hadronically, jets were reconstructed in the hadronic final state using the $k_{T}$ cluster 
algorithm\cite{np:b406:187} in the longitudinally invariant inclusive mode\cite{pr:d48:3160}.
The algorithm was applied to the energy clusters in the CAL after excluding those associated with the scattered-electron 
candidate\cite{epj:c11:427,thesis:briskin:1998,Wai:1995cx}. 
Energy corrections~\cite{pl:b547:164,pl:b558:41,pl:b531:9}  were applied to the jets in order to compensate for energy 
losses in the inactive material in front of the CAL.

In this analysis, only jets with 
$E_{T} > \unit{4\,}{\Gev}$ and $\left| \eta \right| < 2.0$ were used. 
Here  $E_{T}$ is the jet transverse energy and $\eta$ its pseudorapidity.
The hadronic $Z^{0}$ decay sample was selected by the following requirements on the reconstructed jets:
\begin{itemize}
\item at least two jets in the event had to satisfy $E_{T} > \unit{25\,}{\Gev}$;
\item $\left| \Delta \phi_{j} \right| > \unit{2}{\rad}$, where $\Delta \phi_{j}$ is the azimuthal difference between the first and second 
highest-$E_{T}$ jet, as the two leading jets from the $Z^{0}$ boson decays are expected to be nearly back-to-back in the  $X$--$Y$ plane.
\end{itemize}

Electrons were reconstructed using an algorithm that combined information from clusters of energy deposits in the CAL and from tracks\cite{epj:c11:427}. 
To be defined as well-reconstructed electrons, the candidates were required to satisfy the following selection:
\begin{itemize}
\item $ E^{'}_e > \unit{5\,}{\Gev}$ and $ E_{{\rm in}} < \unit{3\,}{\Gev}$, where $ E^{'}_{e}$ is the scattered electron energy and
$ E_{{\rm in}}$ is the total energy in all CAL cells not associated with the cluster of the electron but lying within a cone 
in $\eta$ and $\phi$ of radius $R = \sqrt{ \Delta \eta^{2} + \Delta \phi^{2} } = 0.8$, centred on the cluster;
\item If the electron was in the acceptance region of the tracking system,
a matched track was required with momentum $p_{{\rm track}} > \unit{3\,}{\Gev}$.
After extrapolating the track to the CAL surface, its distance of
closest approach (DCA) to the electron cluster had to be within 10 cm. 
\end{itemize}

The following cuts were applied to suppress low-$Q^{2}$ neutral-current and direct-photopro-\\duction backgrounds:
\begin{itemize}
\item $E_{{\rm RCAL}} < \unit{2\,}{\Gev}$, where $E_{{\rm RCAL}}$ is the total energy deposit in RCAL;
\item $50<E-p_{Z}<\unit{64\,}{\Gev}$,
 where $E-p_{Z} = \sum_{i} E_{i} \left( 1-\cos\theta_{i} \right)$;
$E_{i}$ is the energy of the $i$-th CAL cell, $\theta_{i}$ is its polar angle and the sum runs over all 
cells\footnote{For fully contained events, or events in which the particles escape only in the forward beam pipe, the $E-p_{Z}$ value peaks around twice
the electron beam energy, 55 GeV.};
\item $\theta_{e} < 80^{\circ}$ for well reconstructed electrons, where $\theta_{e}$ is the polar angle of the scattered electron,
motivated by the fact that, due to the large mass of the produced system, the electron is backscattered to the forward calorimeter or forward beam pipe;
\item the event was rejected if more than one electron candidate was found;
\item jets were regarded as a misidentified electron or photon and were discarded from the list of jets if the direction of the jet 
candidate was matched within $R < 1.0$ with that of an electron candidate identified by looser criteria\footnote{
Candidates were selected by less stringent requirements and clusters with no tracks were also accepted to find photons and electrons.}
than those described above.
This cut causes a loss of acceptance of about 3\%.
\end{itemize}

To remove cosmic and beam--gas backgrounds, events fulfilling any of the conditions listed below were rejected:
\begin{itemize}
\item $|Z_{{\rm vtx}}| > \unit{50}{\cm} $, where $Z_{{\rm vtx}}$ is the $Z$ position of the primary vertex reconstructed from CTD+MVD tracks;
\item $175^{\circ} < \left( \theta_{{\rm jet}1} + \theta_{{\rm jet}2} \right) < 185^{\circ}$ and $\Delta \phi_{j} > 175^{\circ}$ simultaneously,
 where $\theta_{{\rm jet}1}$ and $\theta_{{\rm jet}2}$ are the polar angles of the first and second highest-$E_{T}$  jet, respectively, 
and $\Delta \phi_{j}$ is the azimuthal difference between them;
\item $\left| t_{{u}} - t_{d} \right| > \unit{6.0}{\ns}$, where $\left| t_{u} - t_{d} \right|$ is  
the timing difference between the upper and the lower halves of the BCAL;
\item $p\hspace{-.47em}/_{T} > \unit{25\,}{\Gev}$, where $p\hspace{-.47em}/_{T}$ is the missing transverse momentum calculated from the energy clusters in the CAL;
\item $N_{{\rm trk}}^{{\rm vtx}} < 0.25 \left( N_{{\rm trk}}^{{\rm all}} - 20 \right)$, where $N_{{\rm trk}}^{{\rm vtx}}$ is
the number of tracks associated with the primary vertex and $N_{{\rm trk}}^{{\rm all}}$ is the total number of tracks\cite{pl:b539:197}.
\end{itemize}

The number of events passing the above selection was 5257. 
Finally, to select the elastic and quasi-elastic processes preferentially, a cut on $\eta_{{\rm max}}$ was introduced,
\begin{itemize}
\item $\eta_{{\rm max}} < 3.0$.
\end{itemize}
The quantity $\eta_{{\rm max}}$ was defined as the pseudorapidity of the energy deposit in the calorimeter closest to the 
proton beam direction with energy greater than $\unit{400\,}{\Mev}$ as determined by calorimeter cells.
This cut also rejected signal events which have energy deposits from the scattered
electron in the calorimeter around the forward beam pipe, causing an acceptance loss of about 30\%.

After all selection cuts, 54 events remained.
The total selection efficiency was estimated by the MC simulation to be $22\%$ for elastic and quasi-elastic processes and less than $1\%$ for DIS and resolved photoproduction events.
The number of expected signal events in the final sample, as predicted by EPVEC,  is 18.3.
The contribution from elastic and quasi-elastic processes amounts to 17.9 events.

\section{Background-shape study}
\label{bgshape}
\begin{figure}[p]
\vfill
\begin{center}
\includegraphics[width=\hsize]{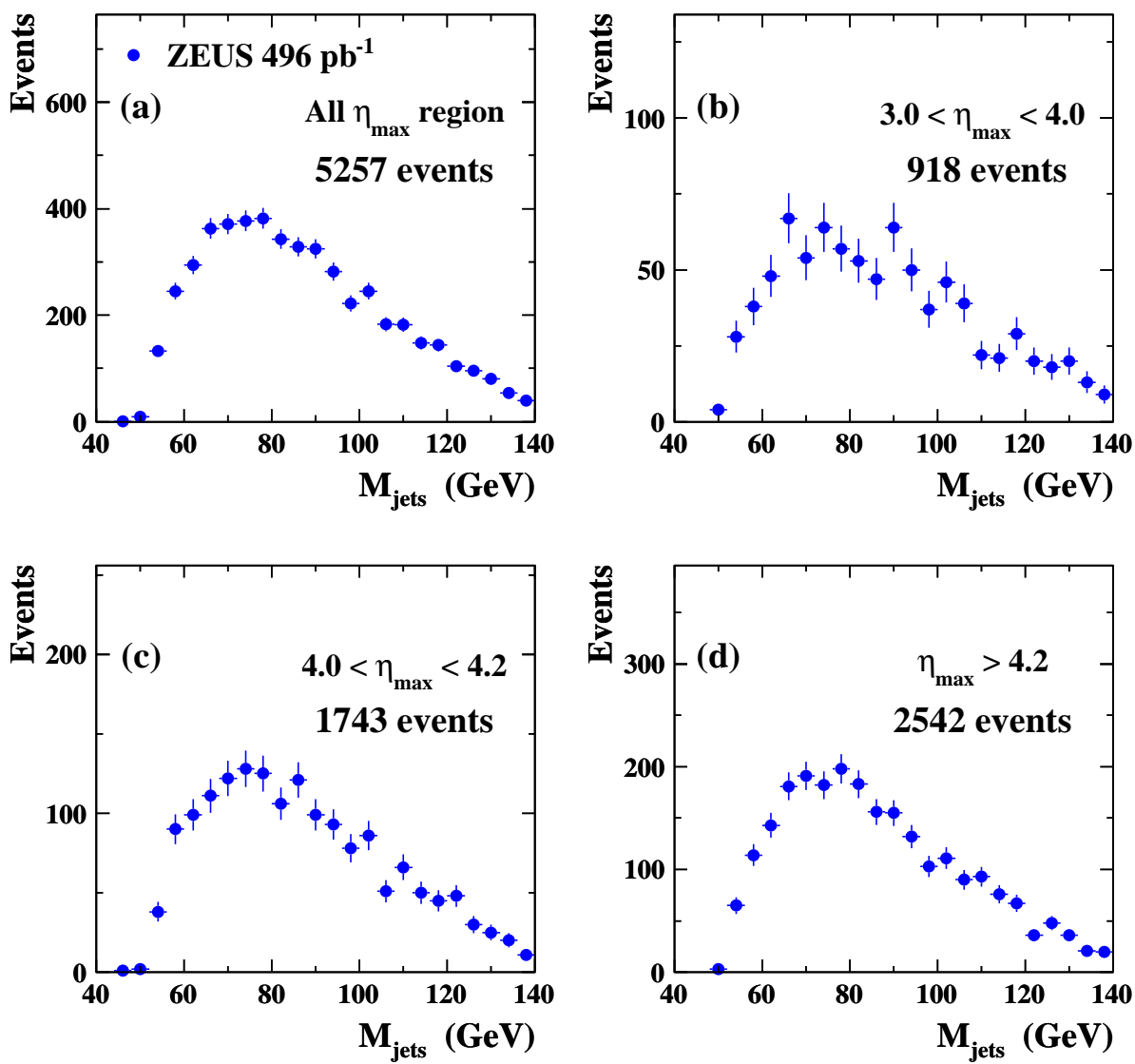}
\end{center}
\caption{The $M_{{\rm jets}}$ distribution of the data (a) after all selection criteria, except for the $\eta_{{\rm max}}$ cut,  
(b-d) in several $\eta_{{\rm max}}$ slices.}
\label{dataall}
\vfill
\end{figure}

\begin{figure}[p]
\vfill
  \begin{center}
\includegraphics[width=0.65\hsize]{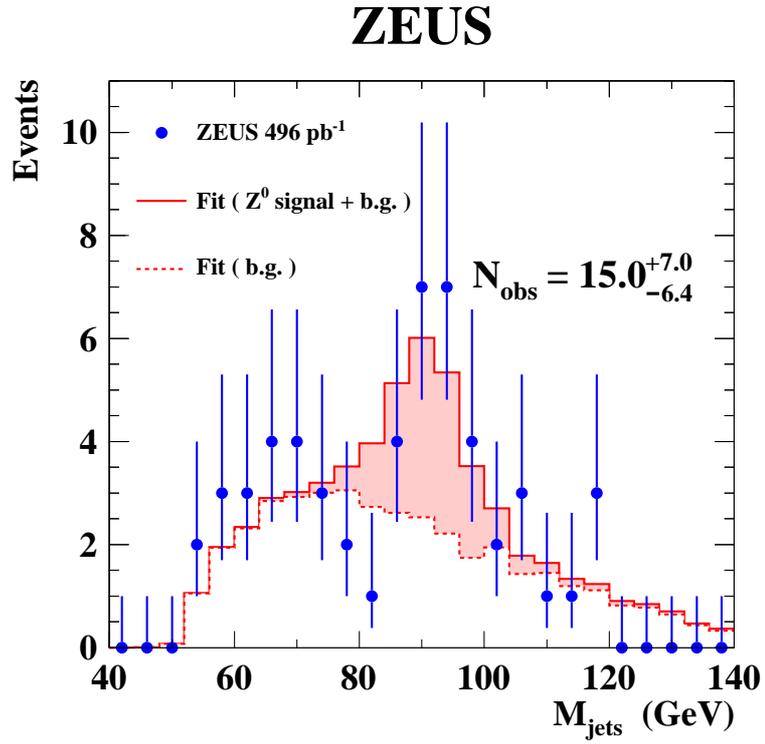}
    \caption{The $M_{{\rm jets}}$ distribution and the fit result. The data are shown as points, and the fitting result of signal+background\,(background component) 
is shown as solid\,(dashed) line. 
The signal contribution is also indicated by the shaded area and amounts
to a total number of $N_{obs}$ events.
The error bars represent the approximate Poissonian 68\% CL intervals, calculated as $\pm \sqrt{n+0.25} + 0.5$ for a given entry $n$.}
    \label{mass_final}
  \end{center}
\vfill
\end{figure}

Figure \ref{dataall}a shows the distribution of the invariant mass, 
$M_{{\rm jets}}$, 
after all the selection criteria except for the requirement $\eta_{{\rm max}} < 3.0$.
The variable $M_{{\rm jets}}$ was calculated using all jets passing the selection criteria described in Section\,\ref{obj_recon}.
Figures \ref{dataall}b-d show $M_{{\rm jets}}$ for various $\eta_{{\rm max}}$ slices 
in the inelastic region\,($\eta_{{\rm max}} > 3.0$) for the same selection.
No significant dependence on $\eta_{{\rm max}}$ of the $M_{{\rm jets}}$ distribution beyond that expected from statistical 
fluctuations was observed in the inelastic region.
In addition, the shape of the $M_{{\rm jets}}$ distribution outside the $Z^{0}$ mass window in the region $\eta_{{\rm max}} < 3.0$ 
was found to be consistent with that in the inelastic region\,(Fig. \ref{mass_final}).
Therefore, the $M_{{\rm jets}}$ distribution in the inelastic region was adopted as a background template in a fit to the data 
in the elastic region as described in the following section.

\section{Cross-section extraction}
A fit to the sum of the signal and a background template for the $M_{{\rm jets}}$ distribution was used for the cross-section extraction.
The template $N_{{\rm ref},i}$ is defined according to:
\begin{equation}
N_{{\rm ref},i} = aN_{{\rm sg},i}^{{\rm MC}}\left( \epsilon \right) + bN_{{\rm bg}, i}^{{\rm data}},
\label{eq1}
\end{equation}
where $i$ is the bin number of the $M_{{\rm jets}}$ distribution. The parameter $\epsilon$ accounts for a possible energy shift, i.e. 
$M_{{\rm jets}}=\left( 1+\epsilon \right)M_{{\rm jets}}^{{\rm MC}}$, where $M_{{\rm jets}}^{{\rm MC}}$ is the invariant-mass distribution 
of the signal $Z^{0}$ MC.
The quantity $N_{{\rm sg},i}^{{\rm MC}}$ is a signal template estimated from the $Z^{0}$ MC distribution after all cuts, 
normalised to data luminosity. 
The quantity  $N_{{\rm bg},i}^{{\rm data}}$ is a background template determined from the data outside the selected region.
The parameters $a$ and $b$ are the normalisation factors for the signal and background, respectively.
The likelihood of the fit, $\mathcal{L}$, is defined as follows:
\begin{equation}
  \mathcal{L} =  \mathcal{L}_{1} \left( N_{{\rm obs}}, N_{{\rm ref}} \right)  \times \mathcal{L}_{2}\left( \epsilon, \sigma_{\epsilon} \right), \\
\label{eq2}
\end{equation}
with
\begin{equation}
 \mathcal{L}_{1} = \prod_{i} \frac{ {\rm exp} \left( -N_{{\rm ref}, i} \right)\left( N_{{\rm ref}, i} \right)^{N_{{\rm obs}, i}} }{ N_{{\rm obs}, i}! } {\rm \ \ \ ~and~ \ \ \ } \mathcal{L}_{2} = {\rm exp}\left( -\frac{\epsilon^{2}}{2\sigma_{\epsilon}^{2}} \right).  \nonumber
\end{equation}
Here $\mathcal{L}_{1}\left( N_{{\rm obs}}, N_{{\rm ref}} \right)$ is the product of Poisson probabilities to observe $N_{{\rm obs}, i}$ events
for the bin $i$ when $N_{{\rm ref}, i}$ is expected.
The term  $\mathcal{L}_{2}\left( \epsilon, \sigma_{\epsilon} \right)$ represents the Gaussian probability density for a shift $\epsilon$ of the jet energy scale from the nominal scale,
which has a known systematic uncertainty of $\sigma_{\epsilon}=3\%$.
From the likelihood, a chi-squared function  is defined as
\begin{equation}
\tilde{ \chi }^{2} = -2\ln \frac{\mathcal{L}_{1}\left( N_{{\rm obs}}, N_{{\rm ref}} \right)}{\mathcal{L}_{1}\left( N_{{\rm obs}}, N_{{\rm obs}} \right)} - 2\ln\mathcal{L}_{2} = 2\sum f_{i}+ \left( \frac{\epsilon}{\sigma_{\epsilon}} \right)^{2}, \\
\end{equation}
with
\begin{equation}
f_{i} =\left\{ \begin{array}{l}
				N_{{\rm ref},i} - N_{{\rm obs},i} + N_{{\rm obs},i}\ln\left( N_{{\rm obs},i}/N_{{\rm ref},i} \right)
				\hspace{11.85mm}\left( {\rm if}\,N_{{\rm obs},i}>0 \right) \\
				N_{{\rm ref},i}  
				\hspace{70mm}\left( {\rm if}\,N_{{\rm obs},i}=0 \right).
				\end{array} 
				\right.\nonumber
\end{equation}
The best combination of ($a$,$b$,$\epsilon$) is found by minimising $\tilde{ \chi }^{2}$.
The value of $a$ after this optimisation gives the ratio between the observed and expected cross section, i.e. 
$\sigma_{{\rm obs}}=a \sigma_{{\rm SM}}$.
The maximum and minimum values of $a$ in the interval $\Delta \tilde{ \chi }^{2} < 1$ define the range of statistical uncertainty.

\section{Systematic uncertainties}
Several sources of systematic uncertainties were considered and their impact on the measurement estimated.
\begin{itemize}
\item An uncertainty of 3\% was assigned to the energy scale of the jets and the effect on the acceptance correction was estimated using the signal MC.
The uncertainty on the $Z^{0}$ cross-section measurement was estimated to be $+2.1\%$ and $-1.7\%$.
\item The uncertainty associated with the elastic and quasi-elastic selection was considered.
In a control sample of diffractive DIS candidate events, the $\eta_{{\rm max}}$ distribution of the MC agreed with the data to within 
a shift of $\eta_{{\rm max}}$ of 0.2 units~\cite{thesis:sola:2012}.
Thus, the $\eta_{{\rm max}}$ threshold was changed in the signal MC by $\pm 0.2$, and variations of the acceptance were calculated accordingly.
The uncertainty on the cross-section measurement was $+6.4\%$ and $-5.4\%$.
\item  The background shape uncertainty was estimated by using different slices of $\eta_{{\rm max}}$ in the fit.
The background shape was obtained using only the regions of $4.0 < \eta_{{\rm max}} < 4.2$ or $4.2 < \eta_{{\rm max}}$.
The region of  $3.0 < \eta_{{\rm max}} < 4.0$ was not used since a small number of signal events is expected in this $\eta_{{\rm max}}$ 
region\footnote{The ratio of the expected number of signal MC events to the observed data in this region was estimated to be 2.6\% 
for $80 < M_{{\rm jets}} < 100$\,GeV, while in the other slices it was less than 0.4\%.}.
The resulting uncertainty in the cross-section measurement was $\pm1.5$\%.
\item The uncertainty associated with the luminosity measurement was estimated to be 2\%, as described in Section \ref{sec-exp}.
\item The $Z^{0}$ mass distribution from the MC used as a signal template has a Gaussian core width of 6~\gev. 
A possible systematic uncertainty coming from the width of the MC signal peak was studied. 
The mass fit was repeated after smearing the $Z^{0}$ mass distribution in the MC by a Gaussian function with different widths. 
The measured cross section did not change significantly after smearing the distribution up to the point where
the fit  $\tilde{ \chi }^{2}$ changed by 1. No systematic uncertainty from this source was assigned.
\end{itemize}
The total systematic uncertainty was calculated by summing the individual uncertainties in quadrature and amounts to $+7.2\%$ and $-6.2\%$.

\section{Results and conclusions}
Figure \ref{mass_final} shows the invariant-mass distribution of the selected events.
It also shows the fit result  
for the signal plus background and the background separately.
The fit yielded a result for the parameter $a$ from Eq.~\ref{eq1} of $a = 0.82^{+0.38}_{-0.35}$.
That translates into a number of observed $Z^{0}$ events of $15.0 ^{+7.0}_{-6.4}$~(stat.), which corresponds
to a signal with a $2.3\,\sigma$ statistical significance.
The fit yielded a value for the energy shift $\epsilon$ of $0.028^{+0.021}_{-0.020}$, which is compatible with zero.
The quality was evaluated according to Eq. 3; the value of $\tilde{ \chi }^{2} / ndf = 17.6 / 22$, 
where $ndf$ is the number of degrees of freedom, indicates a good fit.
The cross section for the elastic and quasi-elastic production of $Z^{0}$ bosons, 
$ep \rightarrow eZ^{0}p^{\left( * \right)}$,  at $\sqrt{s} = \unit{318\,}{\Gev}$, was calculated to be
\begin{equation}
\sigma(ep \rightarrow eZ^{0}p^{\left( * \right)}) = \unit{ 0.13 \pm {0.06} \left( {\rm stat.} \right) \pm{0.01} \left( {\rm syst.} \right) }{\pb}.
\end{equation}
This result is consistent with the SM cross section calculated with EPVEC of $\unit{0.16}{\pb}$.
This represents the first observation of $Z^{0}$ production in $ep$ collisions.


\section*{Acknowledgements}
\label{sec-ack}

\Zacknowledge

\vfill\eject

{
\ifzeusbst
  \bibliographystyle{./l4z_default}
\fi
\ifzdrftbst
  \bibliographystyle{./BiBTeX/bst/l4z_draft}
\fi
\ifzbstepj
  \bibliographystyle{./BiBTeX/user/l4z_epj-ICB}
\fi
\ifzbstnp
  \bibliographystyle{./BiBTeX/bst/l4z_np}
\fi
\ifzbstpl
  \bibliographystyle{./BiBTeX/bst/l4z_pl}
\fi
{\raggedright
\bibliography{
              ./myref.bib,%
              ./l4z_zeus.bib,%
              ./l4z_h1.bib,%
              ./l4z_articles.bib,%
              ./l4z_books.bib,%
              ./l4z_conferences.bib,%
              ./l4z_misc.bib,%
              ./l4z_preprints.bib}}
}
\vfill\eject

%
\end{document}